\documentclass[aps,showpacs,twocolumn]{revtex4} 
\usepackage[dvips]{graphics,graphicx} 
\begin{document}  
\title{Interaction-induced decoherence of atomic Bloch oscillations}  
\author{Andreas Buchleitner} 
\author{Andrey R. Kolovsky} 
\altaffiliation[Also at ]{Kirensky Institute of Physics, 660036 Krasnoyarsk, 
             Russia.} 
\affiliation{Max-Planck-Institut f\"ur Physik komplexer Systeme, D-01187 
             Dresden}   
\date{\today}  
\begin{abstract} 
We show that the energy spectrum of the Bose-Hubbard model   
amended by a static field exhibits Wigner-Dyson    
level statistics. In itself a characteristic signature of quantum chaos,   
this induces the irreversible decay of Bloch oscillations  
of cold, interacting atoms loaded into an optical lattice, and provides a  
Hamiltonian model for interaction induced decoherence.  
\end{abstract}  
\pacs{PACS: 32.80.Pj, 03.65.-w, 03.75.Nt, 71.35.Lk}  
\maketitle 
 
The Bose-Hubbard Hamiltonian serves as a paradigm in the field  
of quantum phase transitions \cite{Sach01}. Recently,  
this model was realized in experiments on ultracold atoms  
loaded into a three-dimensional optical lattice \cite{Grei02},  
opening new perspectives for the laboratory study of correlated 
bosonic systems. Consequently, new theoretical work  
on the Bose-Hubbard model was stimulated, which, in particular,   
addresses the response to a static field  
\cite{Sach02,Brau02,preprint} -- a question  
which shifts the focus from the Bose-Hubbard ground state 
(which is mostly studied in the literature) to dynamical 
and spectral properties of the system. In single-particle quantum  
mechanics, these are associated with Bloch oscillations, in the time domain, 
and with the emergence of a Wannier-Stark  
ladder, in the energy domain 
\cite{PR}.   
     
The present Letter is devoted to the spectral properties    
of the Bose-Hubbard Hamiltonian under the additional action of a      
static field or, equivalently, to the Wannier-Stark problem for   
interacting  bosons. Our analysis is formulated in a spirit close  
to ongoing experiments on cold atoms in optical lattices  
\cite{Grei02,Mors01}, and we assume that the    
atoms are in the ``super-fluid phase'', i.e. they are delocalized over the  
lattice in the absence of any external perturbation.   
This latter assumption distinguishes     
the present work from previous contributions \cite{Sach02,Brau02}   
devoted to the Mott insulator phase, and restricts the values of the     
hopping matrix element $J$ and of the on-site interaction energy $W$ to the  
range $W/J<5.8$ (see \cite{Grei02} and references therein).   
To be specific, we fix $J=0.038$ and $W=0.032$ -- the experimental values  
(in units of photon recoil energies) for rubidium  atoms in optical lattices   
with a potential well depth of approx. ten photon recoils   
-- throughout the sequel of this Letter.  
With $\hat{a}^\dag_{l}$, $\hat{a}_{l}$, and $\hat{n_l}$ the single particle   
creation, the single particle   
anihilation and the number operator at site $l$ of the lattice,    
the total Hamiltonian reads:  
\begin{eqnarray}   
\widehat{H} & = & -\frac{J}{2}\left(\sum_l   
\hat{a}^\dag_{l+1}\hat{a}_l   
+h.c.\right) \nonumber \\  
 & &   
 +\frac{W}{2}\sum_l\hat{n}_l(\hat{n_l}-1)+F\sum_l   
 l\hat{n}_l    
 \; ,  
\label{10}   
\end{eqnarray}   
where the strength of the static field    
$F$ or, more precisely, the Stark energy (the period of lattice is set  
to unity) will be our free parameter.    
\begin{figure}    
\center    
\includegraphics[width=8.5cm, clip]{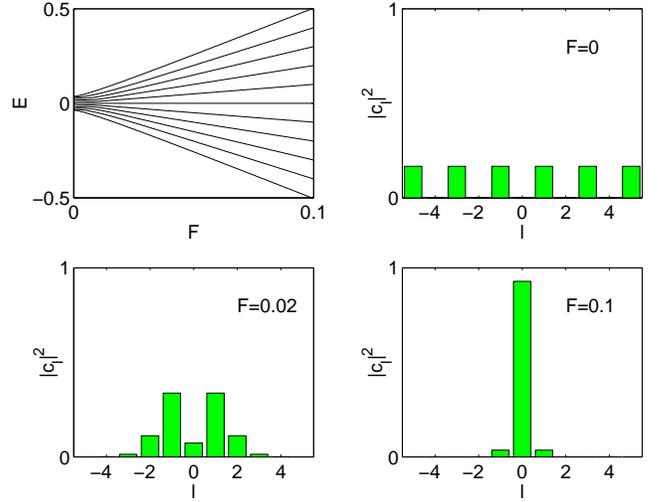}    
\caption{Energy spectrum (top left) together with Wannier state    
projections $|c_l|^2$ of the field-free $k=0$  eigenstate of (\ref{10}),   
for increasing values of the static field $0\leq F\leq 0.1$.     
$N=1$ particle is loaded into a lattice with eleven sites ($L=5$).      
The level dynamics illustrate the transition from   
a Bloch-band to a Wannier ladder, associated with   
the progressive localisation of the wave function in $l$.}     
\label{fig1}    
\end{figure} 
    
Let us first address the issue of boundary conditions. It    
is well known that, for an infinite lattice, there is no    
smooth transition between the spectrum at $F=0$ and $F\ne0$.    
Formally, this is due to the fact that for any non-vanishing value of    
$F$ the Hamiltonian (\ref{10}) is an unbounded operator, whereas    
it is bounded for $F=0$. However, for a lattice    
of finite size, $-L\le l \le L$, the operator (\ref{10}) is always     
bounded and, hence, the spectrum of the system changes continuously     
as a function of $F$, as illustrated by the numerically generated   
level dynamics in the top left panel of Fig.~\ref{fig1},   
for $N=1$ and Dirichlet (i.e., vanishing) boundary conditions.     
to $L-1$). As $F$ is increased,  the spectrum  evolves from 
a Bloch spectrum with energies  
$E(k)=-J\cos[\pi k/(2L+2)]$, $k=-L,\ldots,L$, into a Wannier-Stark ladder   
$E_l\simeq F l$, $l=-L,\ldots,L$. The other panels in Fig.~\ref{fig1}  
show the evolution of the field-free $k=0$ eigenstate in the    
basis of the Wannier states, with increasing $F$ \cite{remark0}.     
The progressive localisation  of the atomic wave function in $l$ with     
$F$ is known as  Stark localisation.    
    
When discussing the   
time evolution of a wave function governed by (\ref{10}),     
it is preferable to use periodic boundary conditions instead of Dirichlet.   
To do so, one first eliminates the static term in (\ref{10}) by transforming   
to the interaction representation, where the hopping and the on-site term   
in (\ref{10})  define the unperturbed Hamiltonian, hence  
\begin{eqnarray}    
\widehat{H}\rightarrow     
\widetilde{H}(t) & = & -\frac{J}{2}\left(\exp\left(-i\frac{F}{\hbar}t\right)  
\sum_l \hat{a}^\dag_{l+1}  \hat{a}_l +h.c.\right) \nonumber \\    
& & +\frac{W}{2}\sum_l \hat{n}_l(\hat{n_l}-1) \;,    
\label{intham}    
\end{eqnarray}  
and then identifies the site $l=L+1$ of the lattice with $l=-L$.    
This choice has the advantage that the time evolution     
operator of a system of non-interacting atoms  over one Bloch period   
$T_B=2\pi\hbar/F$  coincides with the unit matrix,    
independently of the size of the   
system. This facilitates the analysis of the dynamics in the thermodynamic     
limit $N,L\rightarrow\infty$ \cite{preprint}. In what follows, we shall use   
Dirichlet boundary conditions when calculating  eigenvalues, and     
periodic boundary conditions when simulating  the dynamics \cite{remark0+}.  
    
Our analysis of the spectrum of the multi-particle system (\ref{10})        
follows the one for the  single-particle problem. Let us assume for   
the moment that there are no atom-atom interactions, i.e. $W=0$.   
As already mentioned, for large values of $F$ the single-atom energy   
levels form a Wannier ladder, and the energies of an $N$-atom system are    
consequently given by    
\begin{equation}    
\label{11}    
E_{\bf m}=F\sum_{l=-L}^L l m_l\equiv F l_{tot} \;, \quad    
|l_{tot}|\le LN \;,  
\end{equation}    
where the $m_l$ (${\bf m}=m_{-L},\ldots,m_{L}$, $\sum_{-L}^L m_l=N$)     
are the occupation numbers of the Wannier-Stark states.  Note that,   
in general, many different sets $\bf m$ correspond to the same total     
energy, and the $N$-particle Wannier ladder levels $E_{\bf m}=Fl_{tot}$     
are, thus, typically degenerate. The $N$-particle wave function     
associated with a given level $E_{\bf m}$ can be constructed from   
single-particle Wannier-Stark states $|\psi_l\rangle$ by       
an appropriate symmetrisation procedure.  In the basis of Fock states   
(symmetrised products of Wannier functions $|n\rangle$), an arbitrary   
Wannier-Stark state, at finite $F$, is given by the sum    
\begin{equation}    
\label{13}    
|\Psi_{\bf m}\rangle=\sum_{\bf n}c_{\bf n}^{({\bf m})}|{\bf n}\rangle \;,    
\quad |{\bf n}\rangle=|n_{-L},\ldots,n_{L}\rangle \;,    
\end{equation}    
and in the limit $F\rightarrow\infty$ only one coefficient    
$c_{\bf n}^{({\bf m})}$ with ${\bf n}={\bf m}$ differs from     
zero in Eq.~(\ref{13}). On the contrary, in the opposite limit     
$F\rightarrow 0$, almost all expansion coefficients are non-zero and    
the Wannier-Stark states approach $N$-particle Bloch states with     
(once again, degenerate) energies     
\begin{equation}    
\label{14}    
E({\bf k})    
=-J\sum_{k=-L}^L \cos\left(\frac{\pi k}{2L+2}\right) n_k \; ,    
\end{equation}   
the straightforward $N$-particle generalization of the above   
one-particle result.       
\begin{figure}    
\center    
\includegraphics[width=8.5cm, clip]{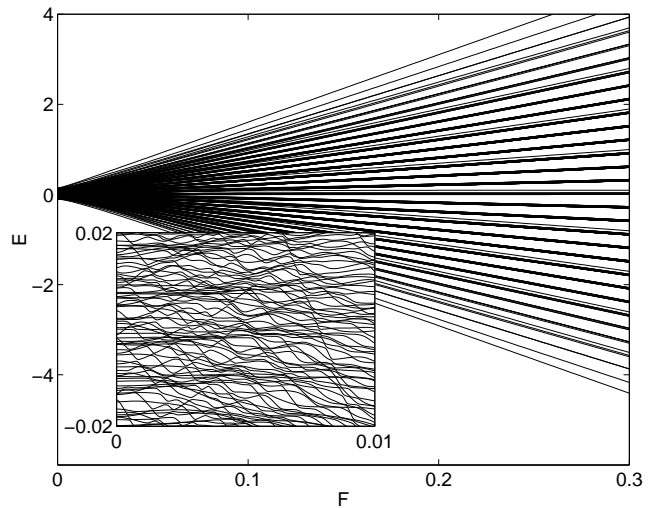}    
\caption{Energy spectrum of the Hamiltonian (\ref{10}) as a    
function of the static field $F$, for particle number     
$N=3$ and lattice size $2L+1=11$. Particle interaction strength $W=0.032$,    
hopping matrix element $J=0.038$. The inset zooms into the central part   
of the spectrum, in the range $0\le F \le 0.01$.}     
\label{fig2}    
\end{figure}    
 
Let us now include the effect of atom-atom interactions.    
Figure~\ref{fig2} shows the energy levels of the Hamiltonian (\ref{10})   
as a function of $F$, for $N=3$ atoms  loaded into a lattice with $11$      
sites (i.e., $L=5$). As expected, the atom-atom interactions    
remove the above-mentioned  degeneracy -- for small $F$ the    
spectrum appears dense (almost continuous), and for large $F$    
the degenerate levels of the Wannier ladder split into     
``Wannier-ladder energy bands''  (see Eq.~(\ref{15}) below).     
In this latter limit, the spectrum and the associated Wannier-Stark   
states can still be found analytically. Indeed, since the hopping     
term in Eq.~(\ref{10}) couples only those Fock states separated by     
one single  quantum in the Stark excitation, one has    
\begin{equation}    
\label{15}    
E_{\bf m}\simeq F\sum_{l=-L}^{L} l m_l  
                +\frac{W}{2}\sum_{l=-L}^{L} m_l(m_l-1) \;,    
\end{equation}
and
\begin{equation}    
\label{16}    
|\Psi_{\bf m} \rangle \simeq |{\bf m}\rangle    
-\frac{J}{2}\left(\sum_{l=-L}^{L-1}\sum_{\bf m'}\frac{\langle {\bf m'}|     
\hat{a}^\dag_{l+1}\hat{a}_l |{\bf m}\rangle}{E_{\bf m'}-E_{\bf m}}     
|{\bf m'}\rangle + h.c.\right) \;,    
\end{equation}    
where $|E_{\bf m'}-E_{\bf m}|\sim F$.   
 
\begin{figure}    
\center    
\includegraphics[width=8.5cm, clip]{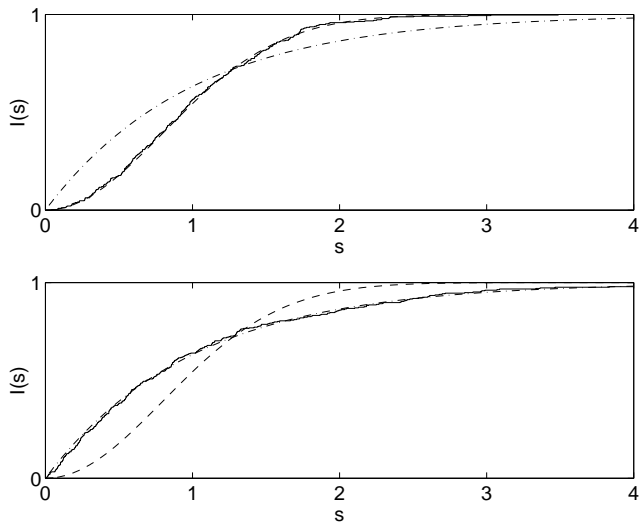}    
\caption{Solid line: Cummulative nearest neighbour level spacing     
distribution $I(s)=\int_0^s P(s')ds'$ for      
normalised spacings $s=\Delta E/\overline{\Delta E}$    
($\overline{\Delta E}$ is the average level spacing in the central part    
of the spectrum), with $F=0.01$ (top) and $F=0.04$ (bottom). $N=4$ atoms   
loaded into a lattice of size $2L+1=11$.   
The dashed and dash-dotted lines indicate GOE and    
Poisson cummulative distributions, respectively.}    
\label{fig3}    
\end{figure}

The perturbative results (\ref{15},\ref{16}) cannot    
hold when $F<J$. Moreover, the complex level dynamics which are borne    
out for small $F$ in the inset in Fig.~\ref{fig2} indicate that    
any attempt to assign a set of quantum numbers to individual levels  
is bound to fail for $F<J$.    
Instead, a statistical analysis of the spectrum is appropriate in this   
situation. For that purpose, the upper part of      
Fig.~\ref{fig3} presents the cummulative distribution of the spacings  
between adjacent energy levels, for $N=4$ atoms loaded into a lattice with    
eleven sites, at $F=0.01$ \cite{remark1}. Clearly,     
the normalised energy intervals $s=\Delta E/\overline{\Delta E}$      
exhibit GOE (Gaussian Orthogonal Ensemble) statistics,     
$P(s)=(\pi^2/6)s\exp(-\pi s^2/4)$, a hallmark of quantum  
chaos \cite{Lesh89}.    
Thus, for weak static fields, the system (\ref{10})    
can be regarded as a quantum chaotic system. The origin of  
``quantum chaos'',    
i.e., of the strongly $F$-dependent, nonperturbative      
mixing of energy levels can be understood here as a consequence     
of the interaction induced lifting of the degeneracy of the    
multi-particle Wannier-Stark levels in the cross-over regime from     
Bloch to Wannier spectra, making nearby levels strongly interact,     
for comparable magnitudes of hopping matrix elements and Stark shifts.    
In contrast, for large $F$ (and in the limit of large $L,N$), the  
nearest neighbour distribution tends towards Poissonian statistics,  
$P(s)=\exp(-s)$, as evident from the lower part of Fig.~\ref{fig3}. 
 
\begin{figure}[!t]    
\center    
\includegraphics[width=8.5cm, clip]{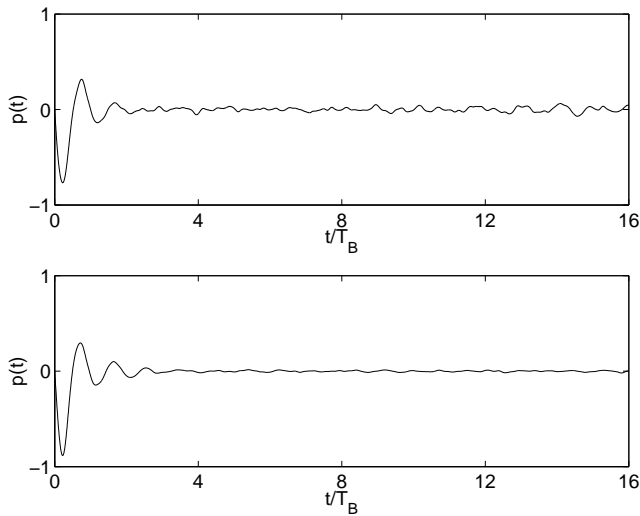}    
\caption{Irreversible decay of atomic Bloch oscillations    
in the presence of a weak static field $F=0.05$, for a filling    
factor $\bar{n}=1$ and lattice sizes $2L+1=7$ (top) and $2L+1=9$    
(bottom). Comparison of both plots suggests that the residual     
fluctuations of $p(t)$ rapidly die out  as the system size is increased.    
(The respective dimensions of Hilbert space are     
${\cal N}=1716$ (top) and ${\cal N}=24310$ (bottom).)}     
\label{fig4}    
\end{figure}

Which are the physical manifestations  of the irregular spectrum of       
(\ref{10}), at intermediate field strengths? To answer this question, we      
consider the Bloch oscillations of the mean atomic  momentum, which can be   
observed rather easily in state of the art  experiments \cite{Mors01,Daha96}.   
In the absence of atom-atom interactions, the average momentum $p(t)$     
of the atoms oscillates with the Bloch frequency $\omega_B=F/\hbar$.         
As shown in \cite{preprint}, the presence of atom-atom interactions     
modifies the Bloch dynamics, and $p(t)$ exhibits an additional beating   
signal at frequency  $\omega_W=W/\hbar$:     
\begin{eqnarray}    
\label{17}    
p(t) &=& {\rm tr}\left[ \hat{p}\hat{\rho}(t)\right ]=NJf(t)\sin(\omega_B t) 
\;,   
\\     
f(t) &=& \exp\left(-2\bar{n}[1-\cos(\omega_W t)]\right) \, ,  \nonumber   
\end{eqnarray}  
with $\bar{n}$ the filling factor, i.e., the average number of     
atoms per lattice cite, and $\hat{\rho}(t)$ the single particle density matrix 
with elements (in the Wannier state basis) 
\begin{equation}  
\label{17a} 
\rho_{l,l'}(t)=\langle\Psi(t)|\hat{a}^\dag_l\hat{a}_{l'}|\Psi(t)\rangle \;.   
\end{equation} 
The appearance of the new frequency $\omega_W$ originates in the 
splitting of the Wannier ladder levels into ``energy bands''  
-- see our above discussion. It must be stressed that the result   
(\ref{17}) is valid     
only for large values of the static field, where the  spectrum is regular.   
Consequently, it is to be expected that for weak static fields      
the atomic Bloch oscillations will be qualitatively different, due to    
the irregular/chaotic structure of the spectrum. Indeed, numerical   
simulations of the dynamics indicate that, in the weak field regime,    
the Bloch oscillations decay irreversibly on rather short time    
scales. As an example, Fig.~\ref{fig4} shows the behaviour of the     
scaled ($p\rightarrow p/NJ$) momentum for  $F=0.05$, $N=7$, $L=3$ (top),   
and $N=9$, $L=4$ (bottom) \cite{remark2}.  After only few Bloch periods, the    
mean momentum has almost comletely decayed but for feeble residual   
fluctuations, which rapidly vanish as the Hilbert space dimension     
is increased (compare top to bottom). Thus, for a weak static field,     
the envelope function in Eq.~(\ref{17}) approaches    
$f(t)\sim\exp(-t/\tau)$ in the thermodynamic limit, which appears    
to be reached rather quickly with increasing system size, by virtue of    
the results shown in Fig.~\ref{fig4}. Note that the decay of the    
Bloch oscillations is due to the decay of the off-diagonal elements    
of the one-particle density matrix (\ref{17a}), and that, hence, the time 
$\tau$ can equally be considered as the {\em decoherence time} for a system of   
interacting bosons. The dependence of $\tau$ on the system parameters     
hitherto remains an open problem.     
      
Finally, let us briefly discuss the conditions for the observed     
chaos-transition  in the Bose-Hubbard model. Our numerical simulations   
of the system dynamics, performed for fixed ratio $W/J$ and    
different values of $N$ and $L$ ($0.2\le\bar{n}\le1.2$,     
$L\leq 10$, $N\leq 10$), suggest the condition    
\begin{equation}    
\label{18}    
\delta l \sim \bar{n}^{-1} \;,     
\end{equation}    
as a criterion of the transition to chaos, where  $\delta l$  denotes the   
localisation length of the  single-particle wave function on the lattice    
($\delta l\simeq J/F$ for $F<J$, and $\delta l\simeq 1$ for $F>J$),   
and $\bar{n}^{-1}$ as the inverse filling factor has the meaning of     
an average particle distance. It is clear, however, that    
condition (\ref{18}) cannot be universal, since it does not     
account for the on-site energy $W$. Indeed, for $W\rightarrow0$,     
the particle-particle interaction vanishes, and the system is integrable for    
arbitrary $F$. On the other hand, when $W\rightarrow\infty$, the    
Bose-Hubbard model can undergo a Mott transition into the insulating phase,   
where its response to the static field has a very different    
(resonant-like) character \cite{Sach02,Brau02}. It therefore     
remains a challenging theoretical problem to formulate general criteria      
for the chaos-transition.  
    
To conclude, we have shown that the spectrum of the Bose-Hubbard   
Hamiltonian amended by a static field (and at fixed particle-particle   
interaction corresponding to the ``superfluid'' regime in field-free case)   
is either regular or irregular, depending on the relative strength of   
the hopping matrix element and the external perturbation.    
In particular, we have seen that the irregular level structure at    
intermediate strengths of the static field manifests in a rapid decay    
of the Bloch oscillations of the mean atomic momentum, and that    
the time scale of this decay provides a direct measure for the decay    
of particle-particle coherences accross the lattice. Hence, chaotic    
dynamics of cold, interacting atoms loaded into a one dimensional    
optical lattice allow for experimental probing and control     
of interaction induced decoherence.   
  
We thank Boris Fine and Henning Schomerus for useful discussions.     
   
    

\begin{thebibliography}{10}
\bibitem{remark0}    
Far from the lattice edges the Wannier-Stark states    
practically coincide with those for an infinite lattice, i.e.    
$|\psi_l\rangle=\sum_n {\cal J}_{l-n}(J/F)|n\rangle$,    
where ${\cal J}_n(z)$ are the ordinary Bessel functions    
and $|n\rangle$ labels the Wannier function associated with    
the $n$th lattice site.
\bibitem{remark0+} Note that the spectral properties determine the dynamcis,  
and that the statistical features of the spectra we are analyzing here are  
robust when we choose periodic instead of vanishing boundary  
conditions. However, the spectral analysis for periodic boundary conditions is  
more involved, since we need to introduce the Floquet-Bloch operator generated  
by $\widetilde{H}(t)$, and will therefore be presented in a separate   
contribution.
\bibitem{remark1}   
For this choice of parameters, the dimension of the Hilbert space and, hence,   
the size of the matrix   
to be diagonalised is  ${\cal N}=(N+2L)!/N!(2L)!=1001$ (to be compared   
to ${\cal N}=286$ for $N=3$ in  Fig.~\ref{fig2} -- we chose $N=4$ here   
to increase the statistical sample). Only the central part of the spectrum,   
with approximately constant density of states, is used for the statistical   
analysis.
\bibitem{remark2}    
Here, the super-fluid state  $|\Psi\rangle\sim     
\left(\sum_l \hat{a}^\dag_l\right)^N |0\ldots0\rangle$    
(which almost coincides with the ground state of the Hamiltonian (\ref{10})   
at $F=0$ and $W/J\sim 1$) is chosen as initial  wave function.   
$F$ is set to a larger value than in Figs.~\ref{fig2} and \ref{fig3}, such    
as to decrease Bloch period $T_B$. Accordingly, $\bar{n}$ is increased such as to   
enhance the observed  decay [what relies on fulfillment of Eq.~(\ref{18})].
\end{thebibliography}
\end{document}